\documentclass[preprint2]{aastex}
\usepackage{graphicx}
\usepackage{amsmath}
\usepackage{amsfonts,amssymb}
\usepackage{times}

\def \vec#1{\ensuremath{\boldsymbol{#1}}}

\def\q{\qquad}

\def \Re {\ensuremath{\rm{Re}}}
\def \Ha {\ensuremath{\rm{Ha}}}
\def \Pm {\ensuremath{\rm{Pm}}}
\def \Rm {\ensuremath{\rm{Rm}}}
\def \Mm {\ensuremath{\rm{Mm}}}
\def \S  {\ensuremath{\rm{S}}}

\def\Om{{\it \Omega}}
\def\A{Alfv\'en}

\renewcommand{\textrm} [1] {\rm #1} 
\renewcommand{\textit} [1] {\it #1}
\renewcommand{\textbf} [1] {\bf #1}

\shorttitle{MHD instabilities of Chandrasekhar flows}
\shortauthors{M. Gellert et al.}

\begin{document}

\title{Nonaxisymmetric MHD instabilities of Chandrasekhar states in Taylor-Couette geometry}

\author{M. Gellert, G. R\"udiger, M. Schultz }
\affil{Leibniz-Institut f\"ur Astrophysik Potsdam, An der Sternwarte 16, D-14482 Potsdam, Germany}
%\email{mgellert@aip.de}

\author{A. Guseva}      
\affil{Institute of Fluid Mechanics, Friedrich-Alexander-Universit\"at Erlangen-N\"urnberg, 91058 Erlangen, Germany}

\and

\author{R. Hollerbach}      
\affil{Department of Applied Mathematics, University of Leeds, Leeds, LS2 9JT, UK}

\begin{abstract}
We consider axially periodic Taylor-Couette geometry with insulating 
boundary conditions. The imposed basic states are so-called Chandrasekhar 
states, where the azimuthal flow $U_\phi$ and magnetic field $B_\phi$ have 
the same radial profiles. Mainly three particular profiles are considered: 
the Rayleigh limit, quasi-Keplerian, and solid-body rotation. In each 
case we begin by computing linear instability curves and their dependence 
on the magnetic Prandtl number $\Pm$. For the azimuthal wavenumber $m=1$ 
modes, the instability curves always scale with the Reynolds number 
and the Hartmann number.  For sufficiently small $\Pm$ these modes 
therefore only become unstable for magnetic Mach numbers less than unity, 
and are thus not relevant for most astrophysical applications. However, 
modes with $m>1$ can behave very differently. For sufficiently flat profiles, 
they scale with the magnetic Reynolds number and the Lundquist number, thereby 
allowing instability also for the large magnetic Mach numbers of astrophysical 
objects.
We further compute fully nonlinear, three-dimensional equilibration of
these instabilities, and investigate how the energy is distributed among the
azimuthal ($m$) and axial ($k$) wavenumbers. In comparison spectra become steeper
for large $m$, reflecting the smoothing action of shear. On the other hand kinetic 
and magnetic energy spectra exhibit similar behavior: if several azimuthal modes 
are already linearly unstable they are relatively flat, but for the rigidly 
rotating case where $m=1$ is the only unstable mode they are so steep that 
neither Kolmogorov nor Iroshnikov-Kraichnan spectra fit the results. The total 
magnetic energy exceeds the kinetic energy only for large magnetic Reynolds 
numbers $\Rm>100$. 
\end{abstract}


%%%%%%%%%%%%%%%%%%%%%%%%%%%%%%%%%%%%%%%%%%%%%%%%%%%%%%%%%%%%%%%%%%%%%%
\section{Introduction}
%%%%%%%%%%%%%%%%%%%%%%%%%%%%%%%%%%%%%%%%%%%%%%%%%%%%%%%%%%%%%%%%%%%%%%%

According to the Rayleigh criterion, an ideal non-magnetic flow is stable
against axisymmetric perturbations whenever the specific angular momentum
increases outward. In the presence of an azimuthal magnetic field $B_\phi$,
this result is modified as
\begin{eqnarray}
 \frac{1}{R^3}\frac{{\rm{d}}}{{\rm{d}}R}(R^2\Om)^2-\frac{R}{\mu_0\rho}
 \frac{{\rm{d}}}{{\rm{d}}R}\left( \frac{B_\phi}{R} \right)^2 > 0,
 \label{mich}
\end{eqnarray}
where $\Om$ is the angular velocity, $\mu_0$ the permeability, $\rho$ the
density, and $(R,\phi,z)$ are standard cylindrical coordinates. This
criterion is both necessary and sufficient for stability against
axisymmetric perturbations \citep{M54}. All ideal flows can thus
be destabilized by adding azimuthal magnetic fields with suitable profiles
and magnitudes.

For nonaxisymmetric modes one has \mbox{${{\rm{d}}}/{{\rm{d}}R} (R B_\phi^2) < 0$}
as the necessary and sufficient condition for stability of an ideal fluid at
rest \citep{V72,T73}. Outwardly increasing fields are
therefore unstable, with azimuthal wavenumber $m=1$ being the most unstable
\citep{A78}. If a differential rotation profile is now added, the variety
of instabilities that are available grows considerably.  Even the current-free
(within the fluid) $B_\phi\propto 1/R$ profile can become unstable, and can
as well be destabilized by a rotation profile that by itself would be stable
according to the Rayleigh criterion. We have called this phenomenon the
Azimuthal MagnetoRotational Instability (AMRI, see \cite{R14});
following theoretical suggestions by \citet{H10}, this mode has
by now been observed in a laboratory experiment \citep{S14}.

This combination of a magnetic field $B_\phi\propto 1/R$ and a rotation profile
$\Om\propto1/R^2$ (potential flow) exactly at the Rayleigh limit is an example of a particular
class of basic states defined by \citet[]{C56} to consist of
\begin{eqnarray}
 \vec{U}=\vec{U}_{\rm A},
 \label{chandra1}
\end{eqnarray}
or more generally,
\begin{eqnarray}
 \vec{U}=\Mm\ \vec{U}_{\rm A}.
 \label{chandra2}
\end{eqnarray}
That is, the radial profiles of $\vec{U}$ and $\vec{U}_{\rm A}=\vec{B}/ \sqrt{\mu_0\rho}$ 
are required to be the same, but there may be a constant of
proportionality between the two, denoted as the magnetic Mach number $\Mm$,
the ratio of the fluid velocity $\vec{U}$ to the \A~velocity $\vec{U}_{\rm A}$
\citep{TM87}. The magnetic Mach number of astrophysical objects
often  exceeds unity. Galaxies have $\Mm$ between 1 and 10 \citep{Els14}, 
for the solar tachocline with a magnetic field of 1 kG one
obtains $\Mm\simeq 30$, and for typical white dwarfs and  neutron stars
$\Mm\simeq 1000$. (On the other hand, for magnetars with fields of
$\sim10^{14}$~G and  a rotation period of $\sim$1 s, the magnetic Mach number
is $\sim0.1-1$.)

\citet[]{C56} showed that all basic states satisfying (\ref{chandra1})
are stable in the absence of diffusive effects. However, these states can be
destabilized if at least one of the molecular diffusivities $\nu$ (kinematic
viscosity) or $\eta$ (magnetic diffusivity) is non-zero. We argued that the
class of states which fulfill the condition (\ref{chandra2}) yield a set of
diffusive instabilities with several properties in common \citep{R15b}. 
While \citet{R15b} concentrated on linear results for the modes $m=\pm1$, this study
extends this work towards higher $m$ and concentrates especially on nonlinear 
effects in the saturated state.
 
As a reminder, for the azimuthal modes $m=1$ the marginal stability curves in
the $\Re$-$\Ha$ plane converge for small magnetic Prandtl numbers 
\begin{eqnarray}\label{def_Pm}
\Pm=\frac{\nu}{\eta}.
\end{eqnarray}
As a consequence, for sufficiently small $\Pm$ instability only exists for
$\Mm<1$, that is, for slow rotation. Rapidly rotating flows with $\Mm>1$
require large $\Pm$ to become unstable. Cosmic objects indeed often possess
small magnetic Prandtl numbers (see \cite{BS05}). For
turbulent systems such as stellar convection zones or galaxies, the magnetic
Prandtl number must be replaced by its effective turbulence-induced values,
which are much larger. In the upper part of the solar radiative core the
molecular value is about $\Pm\simeq 0.065$ \citep{Gou03}. For low-mass red 
giants, however, the inclusion of the radiative viscosity leads to $O(1)$
magnetic Prandtl numbers \citep{R15a}. As many of these magnetized
cosmical objects combine large magnetic Mach numbers with small magnetic
Prandtl numbers, the astrophysical relevance of these Chandrasekhar states,
including AMRI, might seem to be limited. However, these results to date
considered only azimuthal wavenumbers $m=1$. We will see in this work that
$m>1$ modes may behave quite differently, with sufficiently flat profiles
allowing instability for large $\Mm$ even for small $\Pm$, and hence yielding
astrophysically relevant results after all.

Finally, for the sake of completeness, let us return briefly to axisymmetric
modes, and demonstrate that any states satisfying (\ref{chandra2}) are always
stable to such $m=0$ modes, provided only that the rotation rate does not
increase outward. Taking $\Om\propto R^{-q}$ with non-negative $q$, Michael's
relation (\ref{mich}) yields 
\begin{eqnarray}
 (2-q){\Mm}^2 +q>0
 \label{mich2}
\end{eqnarray}
as a sufficient condition for stability. Hence, all flows and fields of the
Chandrasekhar type with $0 \leq q \leq 2$ are stable against axisymmetric
perturbations. Note that the limits $q=0$ and $q=2$ define the two stringent
solutions for the time-independent rotation laws following from the equation of
angular momentum transport. Following \citet{HS06} all rotation
laws between two insulating cylinders under the presence of toroidal fields due
to an axial current inside the inner cylinder are stable against axisymmetric
perturbations. Hence, AMRI in Taylor-Couette flows is strictly nonaxisymmetric.

%%%%%%%%%%%%%%%%%%%%%%%%%%%%%%%%%%%%%%%%%%%%%%%%%%%%%%%%%%%%%%%%%%%%%%%%%%%%%%%%%%%%%%%
\section{Equations}
%%%%%%%%%%%%%%%%%%%%%%%%%%%%%%%%%%%%%%%%%%%%%%%%%%%%%%%%%%%%%%%%%%%%%%%%%%%%%%%%%%%%%%%%
We are interested in the  stability of the background field
$\vec{B}= (0, B_\phi(R), 0)$ and the flow $\vec{U}= (0,R\Om(R), 0)$.
The perturbed state of the system is  described by the field $\vec b$
and the flow $\vec u$.  We will be interested in both linearized and
fully nonlinear solutions to the governing equations. For the
linearized equations all quantities may be expanded in modal form as
${\vec b}={\vec b}(R){\textrm{exp}}(\sigma t+{\textrm{i}}(kz+m\phi)),$
etc., with the axial and azimuthal wavenumbers $k$ and $m$ as `input'
parameters, and $\sigma$ as the (complex) eigenvalue.  The linearized
equations are then
\begin{eqnarray}
 \frac{\partial \vec{u}}{\partial t} + (\vec{U}\cdot\nabla)\vec{u} +  (\vec{u}\cdot\nabla)\vec{U}=
 -\frac{1}{\rho} \nabla p + \nu \Delta \vec{u} + \nonumber\\
 +\frac{1}{\mu_0\rho}{\textrm{curl}}\ \vec{b} \times \vec{B}
 +\frac{1}{\mu_0\rho}{\textrm{curl}}\ \vec{B} \times \vec{b},
\label{mhd}
\end{eqnarray}
\begin{eqnarray}
 \frac{\partial \vec{b}}{\partial t}= {\textrm{curl}} (\vec{u} \times \vec{B})+  {\textrm{curl}} (\vec{U} \times \vec{b})+\eta \Delta\vec{b},
\label{mhd1}
\end{eqnarray}
and
${\textrm{div}}\ \vec{u} = {\textrm{div}}\ \vec{b} = 0$.  For the full
nonlinear problem (\ref{mhd}) contains the additional terms
$(\vec{u}\cdot\nabla)\vec{u}$ on the left and
$({\textrm{curl}}\ \vec{b} \times \vec{b})/(\mu_0\rho)$ on the right,
and (\ref{mhd1}) contains the additional term ${\textrm{curl}}(\vec{u}\times\vec{b})$ 
on the right. The modal expansion above also no longer
holds; the spatial structure is instead allowed to be fully three-dimensional,
and the evolution in time is via time-stepping rather than an eigenvalue
problem.

The  stationary background solutions which fulfill the condition
(\ref{chandra2}) are
\begin{eqnarray}
 \Om=a +\frac{b}{R^2},  \ \ \ \ \ \ \ \ \ \ \ \ \ \ \ \ \ B_\phi= \frac{\sqrt{\mu_0\rho}}{\Mm}\ (a R+\frac{b}{R}),
\label{basic}
\end{eqnarray}
where $a$ and $b$ are constants defined by 
\begin{eqnarray}
 a=\Om_{\rm{in}}\frac{ \mu-r_{\rm in}^2}{1-{r_{\rm in}^2}}, \q
 b=\Om_{\rm{in}} R_{\rm{in}}^2 \frac{1-\mu}{1-{r_{\rm in}^2}},
 \label{ab}
\end{eqnarray}
with
\begin{equation}
 r_{\rm in}=\frac{R_{\rm{in}}}{R_{\rm{out}}}, \; \; \; \ \ \ \ \ \ \ \mu=\frac{\Om_{\rm{out}}}{\Om_{\rm{in}}}.
 \label{mu}
\end{equation}
$R_{\rm{in}}$ and $R_{\rm{out}}$ are the radii of the inner and outer
cylinders, and $\Om_{\rm{in}}$ and $\Om_{\rm{out}}$ are their rotation
rates. A magnetic field of the form $b/R$ is generated by running an
axial current only through the inner region $R<R_{\rm{in}}$, whereas a
field of the form $a R$ is generated by running a uniform axial current
through the entire region $R<R_{\rm{out}}$, including the fluid.  

The toroidal field amplitude is usually measured by the Hartmann number
\begin{eqnarray}
 \Ha = \frac{B_{\rm in} R_0}{\sqrt{\mu_0 \rho \nu \eta}}
 \label{Ha}
\end{eqnarray}
of the azimuthal field  $B_{\rm{in}}$ at the inner cylinder.
$R_0=\sqrt{R_{\rm in}(R_{\rm out} - R_{\rm in})}$ is used as the unit of
length, $\eta/R_0$ as the unit of velocity and $B_{\rm in}$ as the unit
of the azimuthal fields. Frequencies, including the rotation $\Om$, are
normalized with the inner rotation rate $\Om_{\rm in}$. The Reynolds
numbers $\Re$ and $\Rm$ are defined by
\begin{eqnarray}
 \Re=\frac{\Om_{\rm in}  R_0^2}{\nu},
 \ \ \ \ \ \ \ \ \ \ \ \ \ \ {\Rm}=\frac{\Om_{\rm in}  R_0^2}{\eta},
 \label{Re}
\end{eqnarray}
and the magnetic Mach number is then related via
\begin{eqnarray}
  \Mm=\frac{\sqrt{\Re\Rm}}{\Ha}=\frac{\Rm}{\S}
  \label{Mm}
\end{eqnarray}
with the Lundquist number $\S=\Ha \cdot \sqrt{\Pm}$ of the magnetic field.

The boundary conditions imposed at $R_{\rm{in}}$ and $R_{\rm{out}}$
are no-slip for $\vec u$ and insulating for $\vec b$. This translates to
\begin{eqnarray}
 u_R=u_\phi=u_z=0
 \label{ubnd}
\end{eqnarray}
at both boundaries,
\begin{eqnarray}
 b_R+ {{\rm i} b_z \over I_m(kR)} \left({m\over kR} I_m(kR)+I_{m+1}(kR)\right)=0
 \label{BR1}
\end{eqnarray}
at $R_{\textrm{in}}$, and  
\begin{eqnarray}
 b_R+{{\rm i} b_z \over K_m(kR)} \left({m\over kR} K_m(kR) - K_{m+1}(kR)\right)=0
 \label{B_R2}
\end{eqnarray}
at $R_{\rm out}$, where $I_m$ and $K_m$ are the modified Bessel functions.
A more detailed derivation of the boundary conditions can be found in \citet{R13}.

We fixed the radius ratio at $r_{\rm in}=0.5$. For the rotation ratio we then
consider primarily the three values $\mu=0.25$, $1$ and $0.35$. The choice
$\mu=0.25$ corresponds to a flow that is exactly at the Rayleigh limit
$\Om\propto 1/R^2$, and a field that is current-free within the fluid; any
instabilities are therefore pure AMRI. The choice $\mu=1$ corresponds to a
solid-body rotation, and a uniform electric current flowing throughout the
entire region (what is known as a `pinch' configuration in plasma physics).
Any instabilities in this case are purely current-driven, what are also known
as Tayler instabilities (TI).  We will find that $m=1$ are the only
instabilities in this case. The choice $\mu=0.35$ has aspects in common with
both the AMRI and TI; that is, instabilities in this case can derive their
energy from either the background flow $\vec U$ (AMRI) or the background
field $\vec B$ (TI). The reason for the particular choice $\mu=0.35$ is that
this represents the so-called quasi-Keplerian value where
$\Om\approx R^{-3/2}$, although according to (\ref{basic}) the profile is
not exactly Keplerian, but merely has the values of $a$ and $b$ that fit a
Keplerian ratio at the endpoints. Finally, a few calculations were also done
at $\mu=0.5$, which corresponds to a so-called quasi-galactic profile, where
$a$ and $b$ are fitted to $\Om\approx R^{-1}$.

The linearized one-dimensional eigenvalue problem is solved using the
numerical code described by \citet{R13}, as well as further references
therein. The nonlinear three-dimensional time-stepping problem is solved
using the MPI-parallelized code described by \citet{Gus15}, which
itself is based on an earlier pipe flow solver by A.P. Willis
(www.openpipeflow.org). The spatial structures in $z$ and $\phi$ are via
Fourier modes $\exp(ikz+im\phi)$, allowing energy spectra in these two
directions to be easily constructed. The periodic domain length in the 
axial direction is chosen as 10 times the gap width, to allow sufficient 
large structures to develop in $z$. Usually close to the linear onset of the
instability the wavenumbers in axial direction conform to the gap width, thus 
they are well-captured. In axial direction between 64 and 256 Fourier modes 
have been used, in azimuthal between 32 and 128. For the radial direction the order of 
Chebyshev polynomials was varied between 127 and 511. In summary, the lowest
resolution has been $127 \times 64\times 32$, the highest $511 \times 256 \times 128$, 
depending mainly on the magnetic Reynolds number.

In the next section we use the linear code to investigate the
onset of instabilities for our chosen values of $\mu$; in the section after
that we use the nonlinear code to study their equilibration in the
supercritical regime.

%%%%%%%%%%%%%%%%%%%%%%%%%%%%%%%%%%%%%%%%%%%%%%%%%%%%%%%%%%%%%%%%%%%%%%%%%%
\section{Linear Onset}
%%%%%%%%%%%%%%%%
We wish to compute the linear onset curves for the three azimuthal
wavenumbers $m=1,\ 2,\ 3$, and the values $\mu=0.25$, 1, and 0.35 (plus a
few results at 0.5). That is, for each choice of input parameters $\Ha$,
$\Re$ and $\Pm$, we repeatedly solve the linear eigenvalue problem for a
range of $k$, and find the value that yields the largest growth/decay rate,
${\rm Re}(\sigma)$. The curve where ${\rm Re}(\sigma)=0$ is then the linear
onset curve, and we are particularly interested in how this curve scales as
$\Pm\to0$. Are the relevant parameters $\Ha$ and $\Re$, or $\S$ and $\Rm$,
and does this perhaps differ for different values of $m$ and $\mu$?

%%%%%%%%%%%%%%%%%%%%%%%%%%%%%%%%%%%%%%%%%%%%%%%%%%%%%%%%%%%%
\subsection{The Rayleigh limit, $\boldsymbol{\mu=0.25}$}
%%%%%%%%%%%%%%%%%%%%%%%%%%%%%%%%%%%%%%%%%%%%%%%%%%%%%%%%%%%%%%%
\begin{figure*}[htb]
\centering
\includegraphics[width=0.32\textwidth]{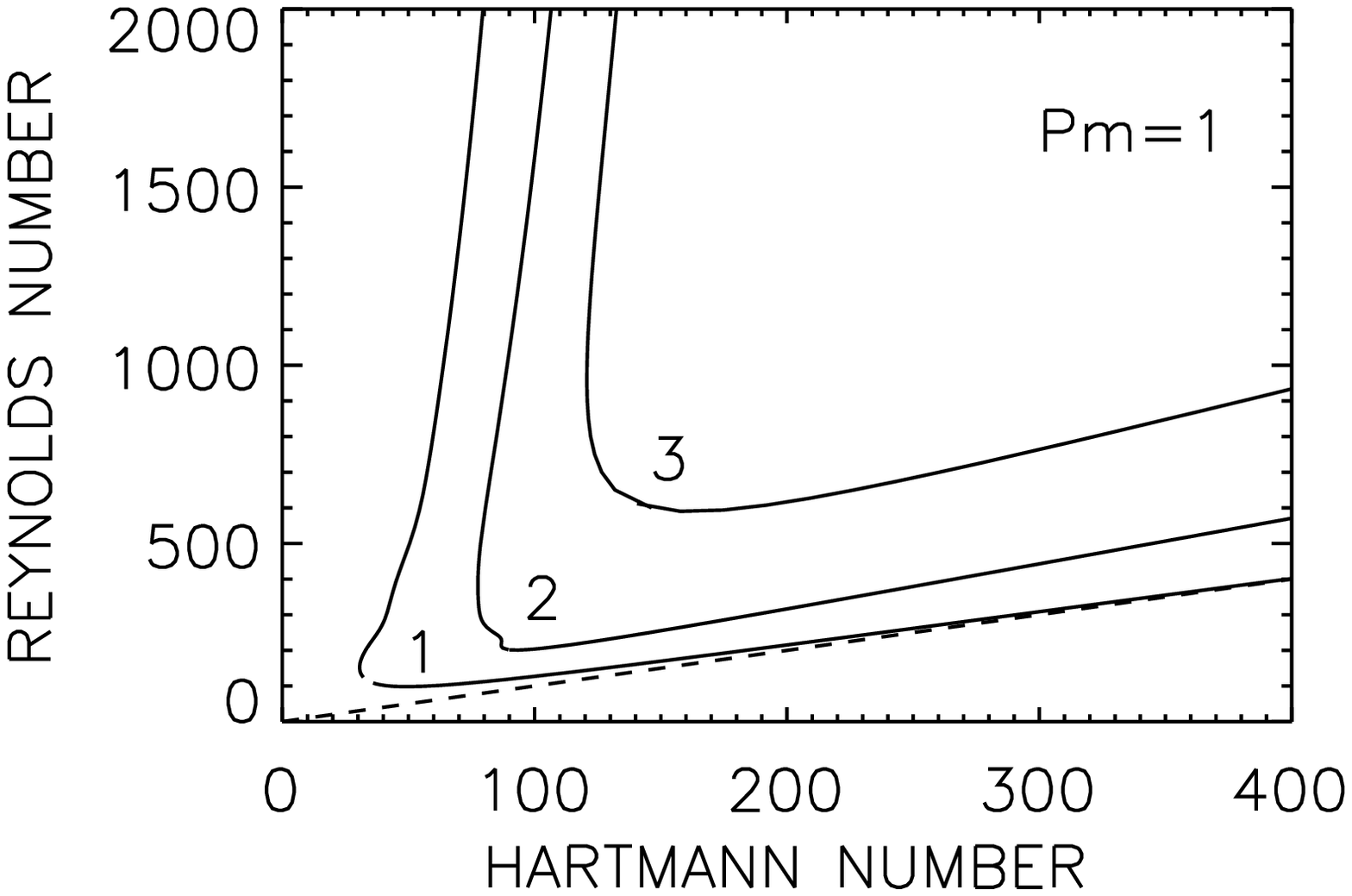}
\includegraphics[width=0.32\textwidth]{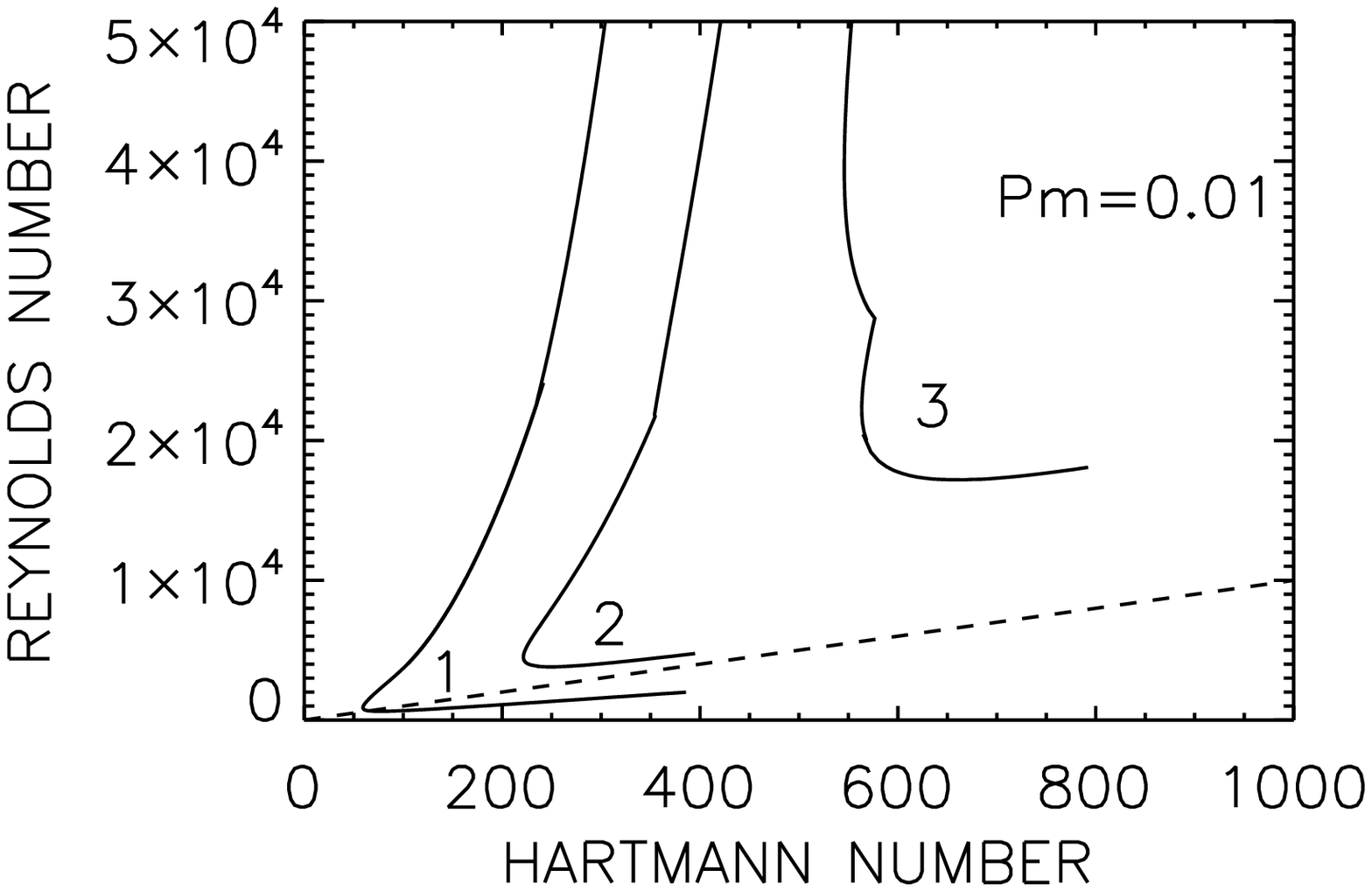}
\includegraphics[width=0.32\textwidth]{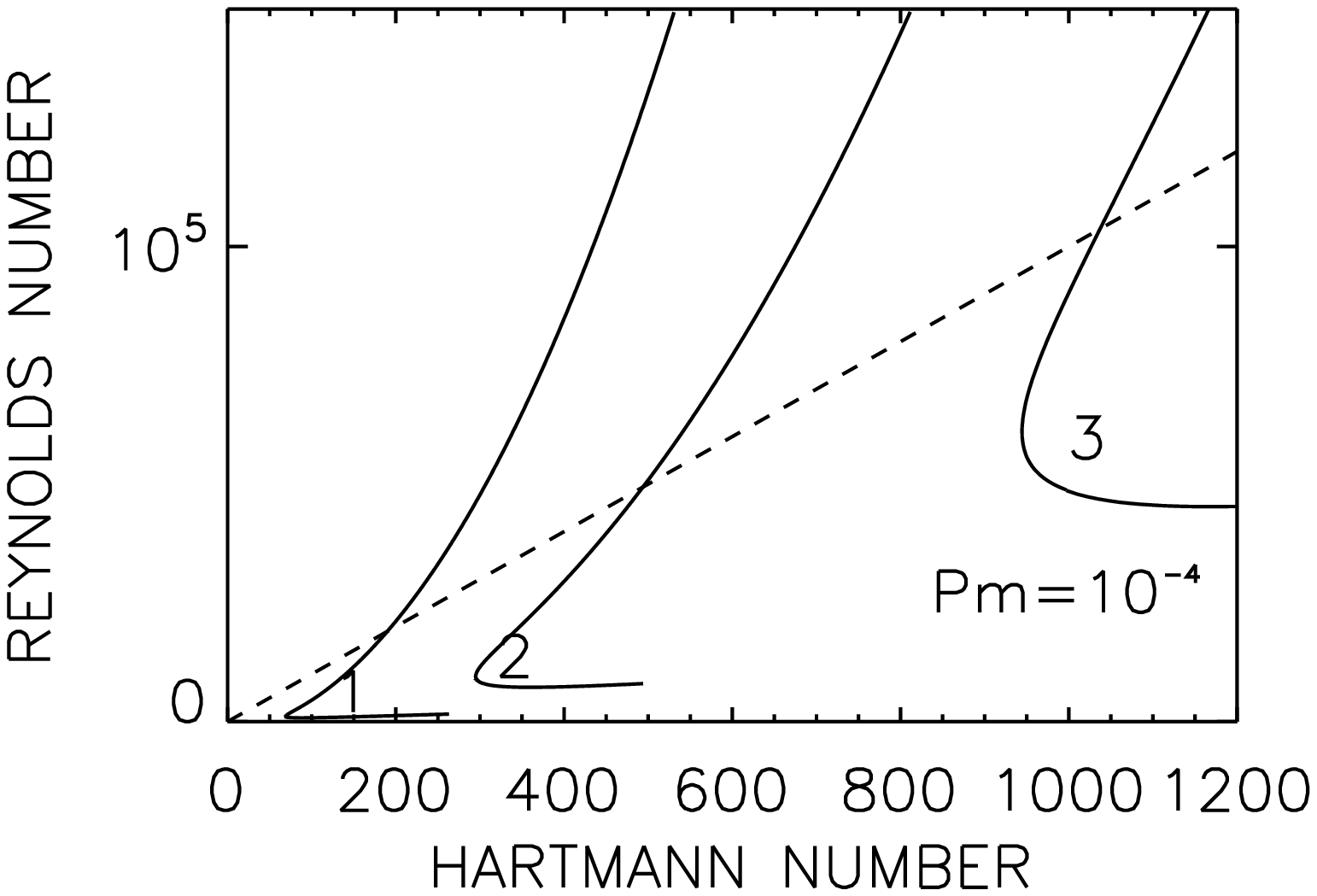}
\caption{The stability maps for $\mu=0.25$ for $m=1,2,3$ and $\Pm=1$ (left),
         $\Pm=10^{-2}$ (middle) and $\Pm=10^{-4}$ (right). The dashed lines
         define $\Mm=1$. For $\Pm\to 0$ all curves satisfy the condition $\Mm<1$.}
\label{25a}
\end{figure*}

The value $\mu=0.25$ has a particular significance for both the flow and the
field. For $\vec U$, it denotes the transition point from hydrodynamic
instability for $\mu<0.25$ to stability for $\mu>0.25$, according to the
Rayleigh criterion regarding the angular momentum $R^2\Om$. For $\vec B$ we
have that the associated electric currents flow only in the inner region
$R<R_{\textrm{in}}$. Any resulting instabilities are therefore purely
magnetorotational in nature, not current-driven. As a result, no
instabilities can occur for $\Re=0$; $\Ha=0$ is also excluded, as $\mu=0.25$
is already on the Rayleigh line where purely non-magnetic instabilities no
longer exist.

\begin{figure}[htb]
\centering
\includegraphics[width=0.38\textwidth]{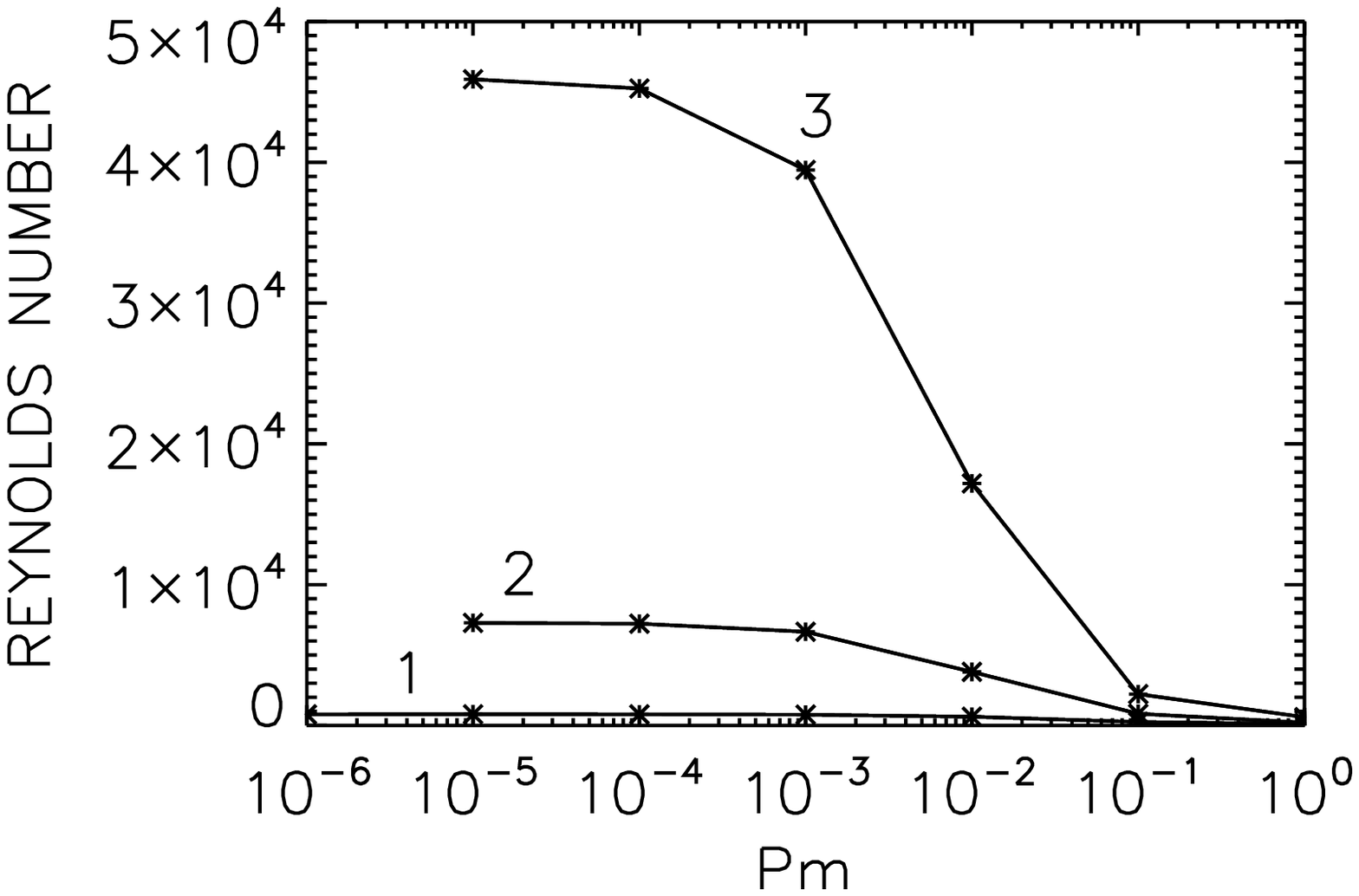}
\includegraphics[width=0.38\textwidth]{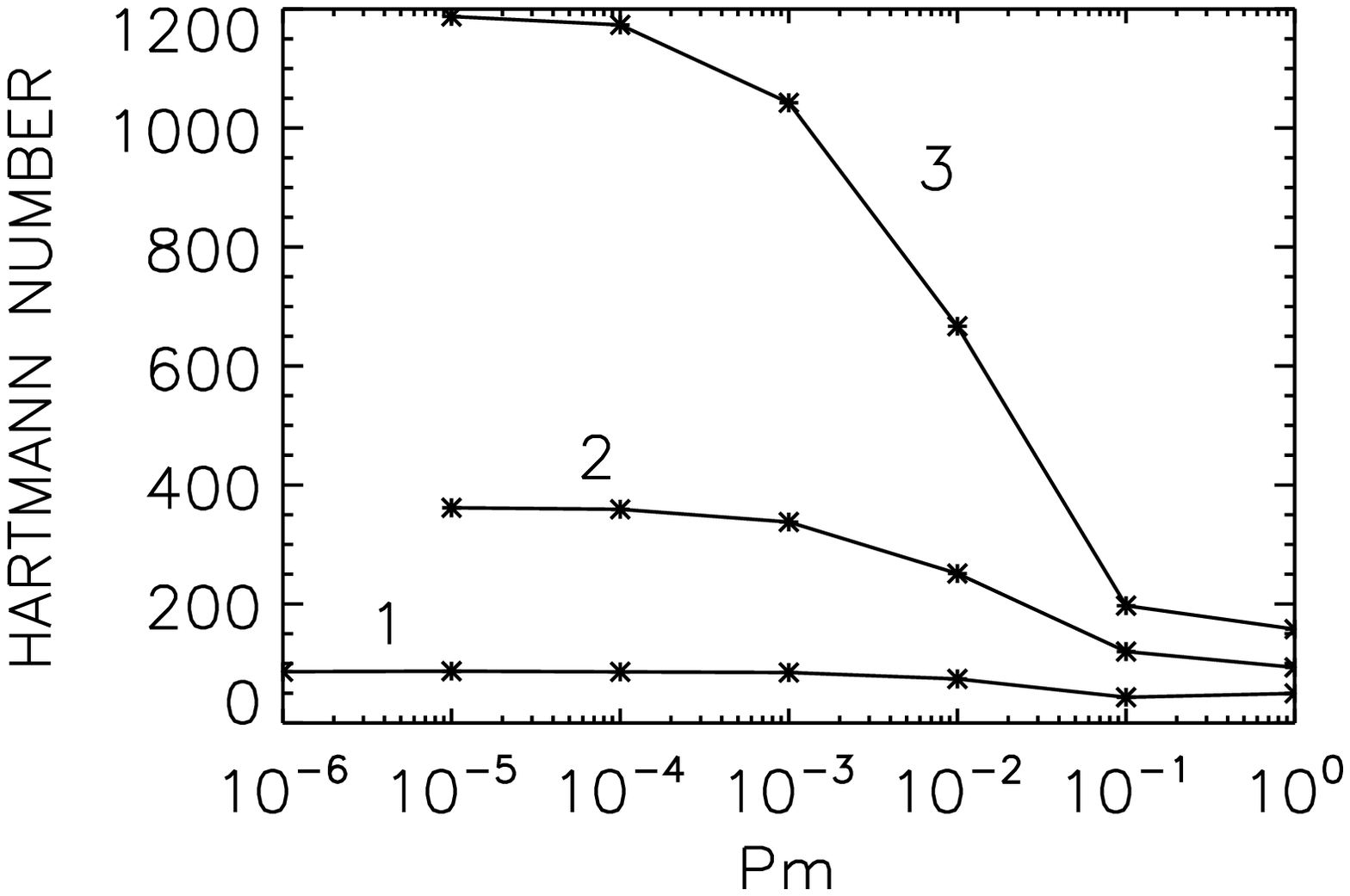}
\caption{The coordinates $\Re$ (top) and $\Ha$ (bottom) for the minima  of the lines
         of marginal instability given in Fig. \ref{25a}. For small $\Pm$ the lines for
         {\em all} $m$ scale with $\Re$ and $\Ha$.}
\label{25b}
\end{figure}

Fig. \ref{25a} shows results for $\Pm=1$ to $\Pm=10^{-4}$. For all $m$, the
curves have a characteristic shape consisting of lower and upper branches
that each have positive slopes. That is, for a sufficiently large $\Ha$ to
allow instability at all, it only exists within a finite range $\Re_l\le
\Re\le\Re_u$, and vice versa when interchanging the roles of $\Ha$ and $\Re$.
The global minimum values of $\Re$ and $\Ha$ are plotted in Fig. \ref{25b}.
Figs. \ref{25a} and \ref{25b} clearly reveal that: (i) The modes $m=2$ and 3
are also unstable, but $m=1$ is always the most unstable; (ii) Decreasing
$\Pm$ pushes the onset to higher values of $\Re$ and $\Ha$, and more
strongly for $m=2$ and 3 than for $m=1$; (iii) For sufficiently small $\Pm$
the critical parameters for all three azimuthal modes are $\Re$ and $\Ha$.
This last result in particular means that as $\Pm\to0$ all of the onset curves
shift increasingly into the regime $\Mm<1$, making them astrophysically not
relevant. On the other hand, it is precisely this feature that the scalings
are $\Re$ and $\Ha$ rather than $\Rm$ and $\S$ that made these modes
experimentally accessible \citep{H10,S14}.

%%%%%%%%%%%%%%%%%%%%%%
\subsection{Rigidly-rotating pinch, $\boldsymbol{\mu=1}$}\label{subsec_pinch}
%%%%%%%%%%%%%%%%%%%%%

The value $\mu=1$ also has special significance for both the flow and the
field. For $\vec B$ it implies a uniform current throughout the entire
region $R<R_{\textrm{out}}$, what is known in plasma physics as a pinch
configuration. For $\vec U$, it corresponds to solid-body rotation, with no
differential rotation at all. Any resulting instabilities are therefore
purely current-driven, with $\vec U$ not available as a source of energy.
As a result, instabilities can occur for $\Re=0$ (corresponding to a
stationary container), but not for $\Ha=0$.

\begin{figure}[htb]
\centering
\includegraphics[width=0.38\textwidth]{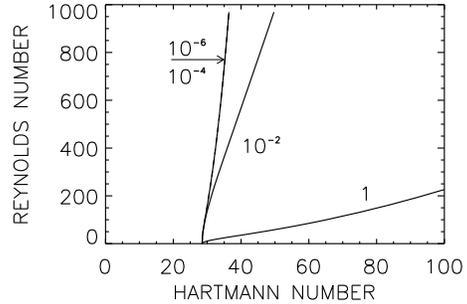}
\caption{The stability maps for $\mu=1$, $m=1$ (the only unstable mode), and $\Pm$
         as indicated next to each curve. Note how the curves become identical for
         $\Pm\le10^{-4}$.}
\label{100a}
\end{figure}

Fig. \ref{100a} shows results for $\Pm=1$ to $\Pm=10^{-6}$. All curves start
at $\Ha=28.1$ for $\Re=0$, then curve toward the right for $\Re>0$. That
is, solid-body rotation has a stabilizing influence, which is strongest for
$\Pm=1$ \citep{PT85}. Note also that only $m=1$ is unstable in this
case. For $\Re=0$ this was previously known \citep{T57}; we here extend
this result to $\Re>0$. The other key message from Fig. \ref{100a} is that
once again, for sufficiently small $\Pm$ the critical parameters are $\Re$
and $\Ha$, so $\Mm<1$. And again, it is precisely this feature that is
experimentally so convenient \citep{R07, R10, S12}.

%%%%%%%%%%%%%%%%%%%%%
\subsection{Quasi-Keplerian rotation, $\boldsymbol{\mu=0.35}$}
%%%%%%%%%%%%%%%%%%%%%

The previous results at $\mu=0.25$ and $1$ have been particularly simple,
in the sense that any instabilities are necessarily either pure AMRI or
pure TI, based simply on the energy source that is driving the instability.
Any values in between, including the astrophysically relevant
quasi-Keplerian profile $\mu=0.35$, or also the quasi-galactic $\mu=0.5$,
are potentially far more complicated, as both $\vec U$ and $\vec B$ can
act as energy sources. Not surprisingly then, the results are also more
complicated than either of the `pure' cases.

\begin{figure*}[htb]
\centering
\includegraphics[width=0.32\textwidth]{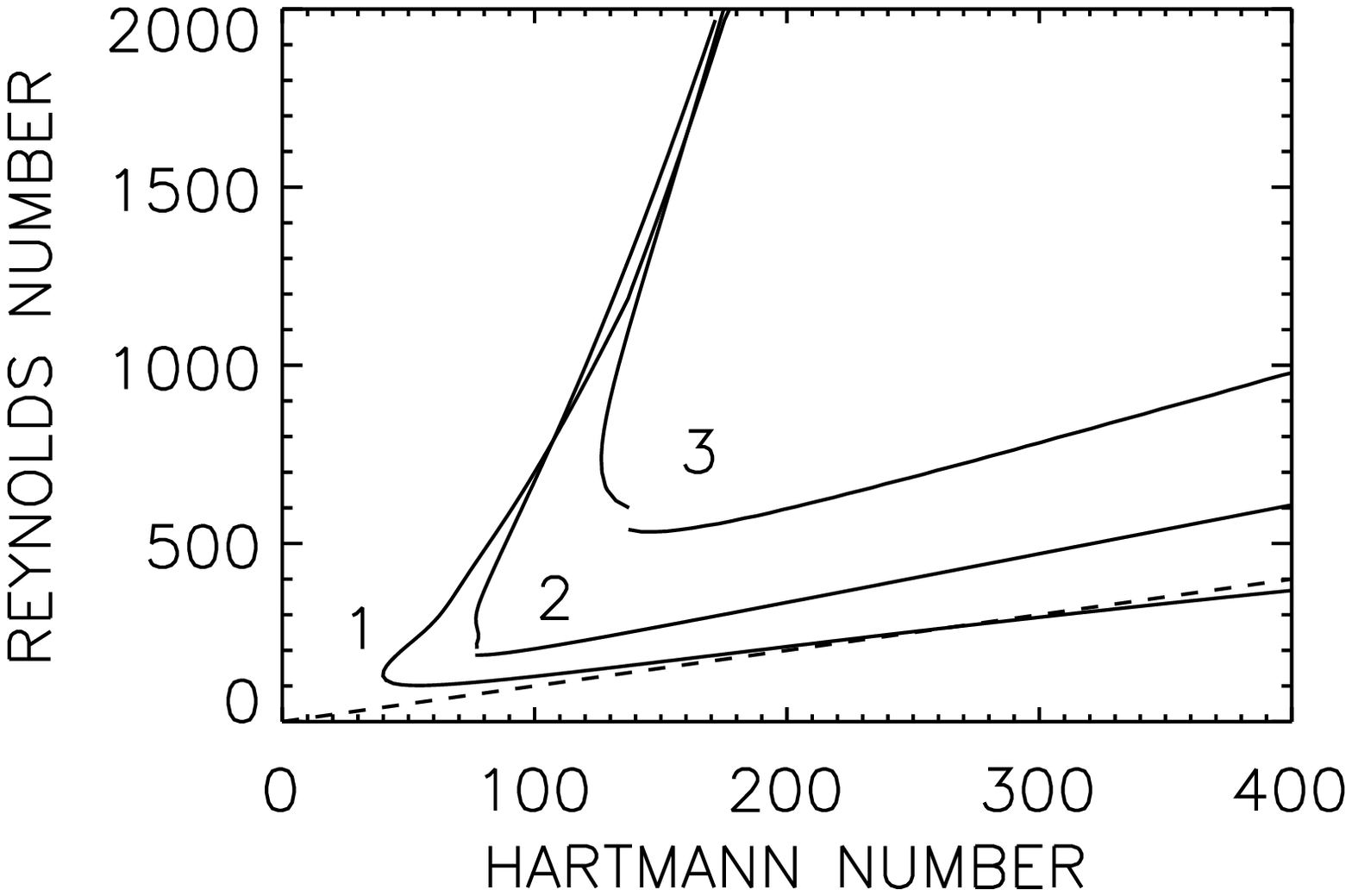}
\includegraphics[width=0.32\textwidth]{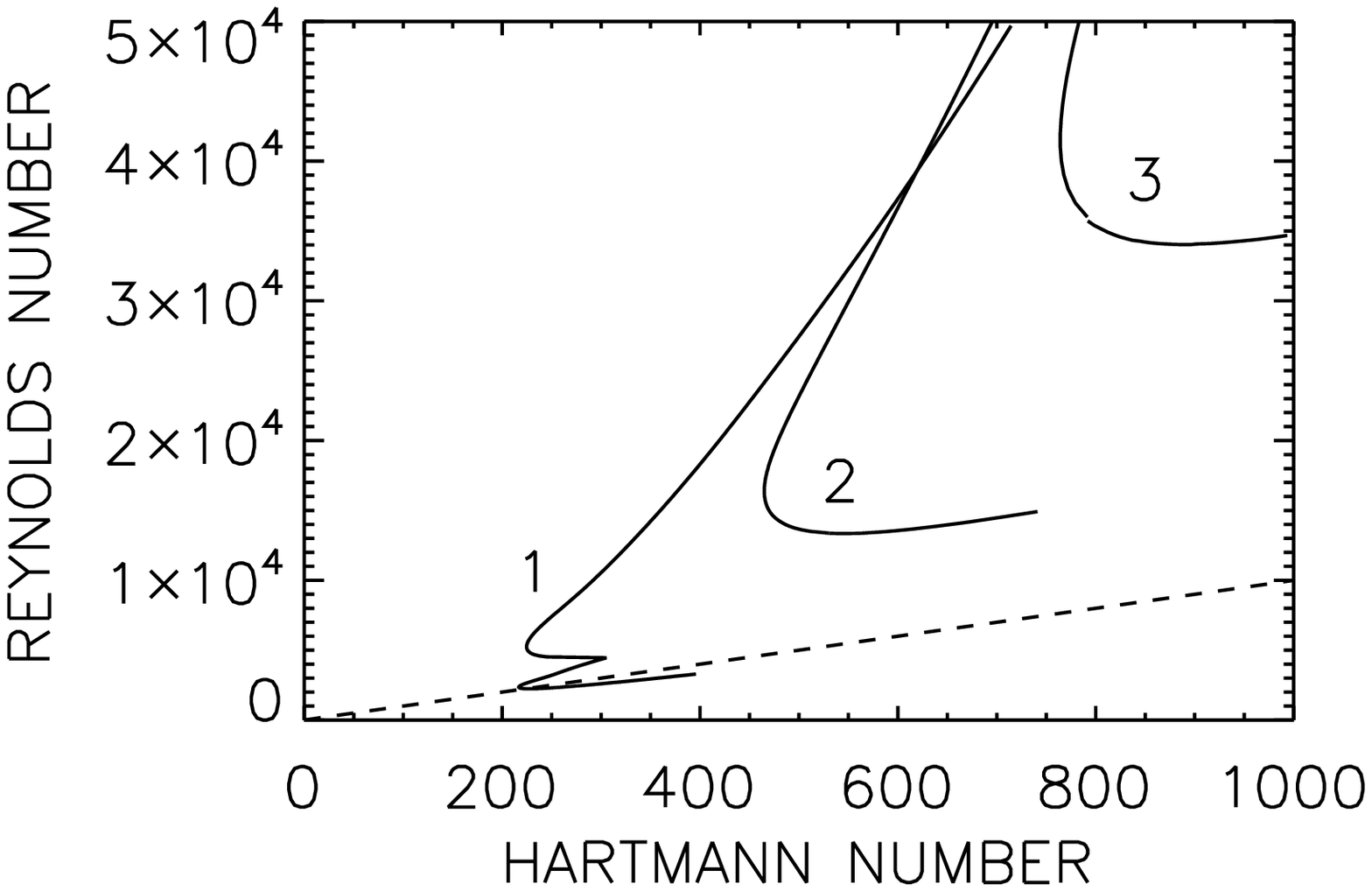}
\includegraphics[width=0.32\textwidth]{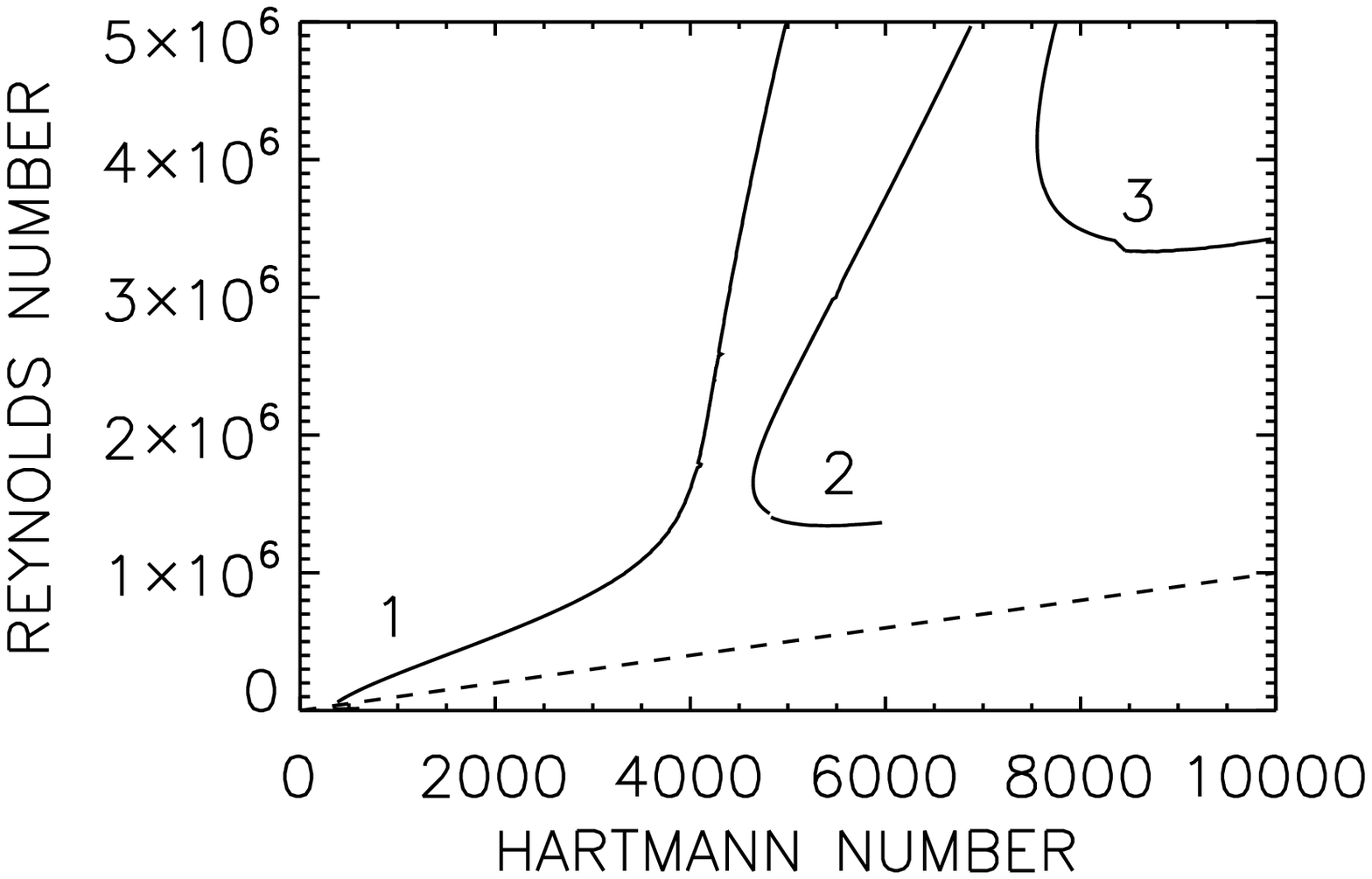}
\caption{The stability maps for $\mu=0.35$ for $m=1,2,3$ and $\Pm=1$ (left),
         $\Pm=10^{-2}$ (middle) and $\Pm=10^{-4}$ (right). The dashed lines
         define $\Mm=1$. For $\Pm\to 0$ only the $m=1$ curve satisfies the
         condition $\Mm<1$.}
\label{35a}
\end{figure*}

\begin{figure}[htb]
\includegraphics[width=0.22\textwidth]{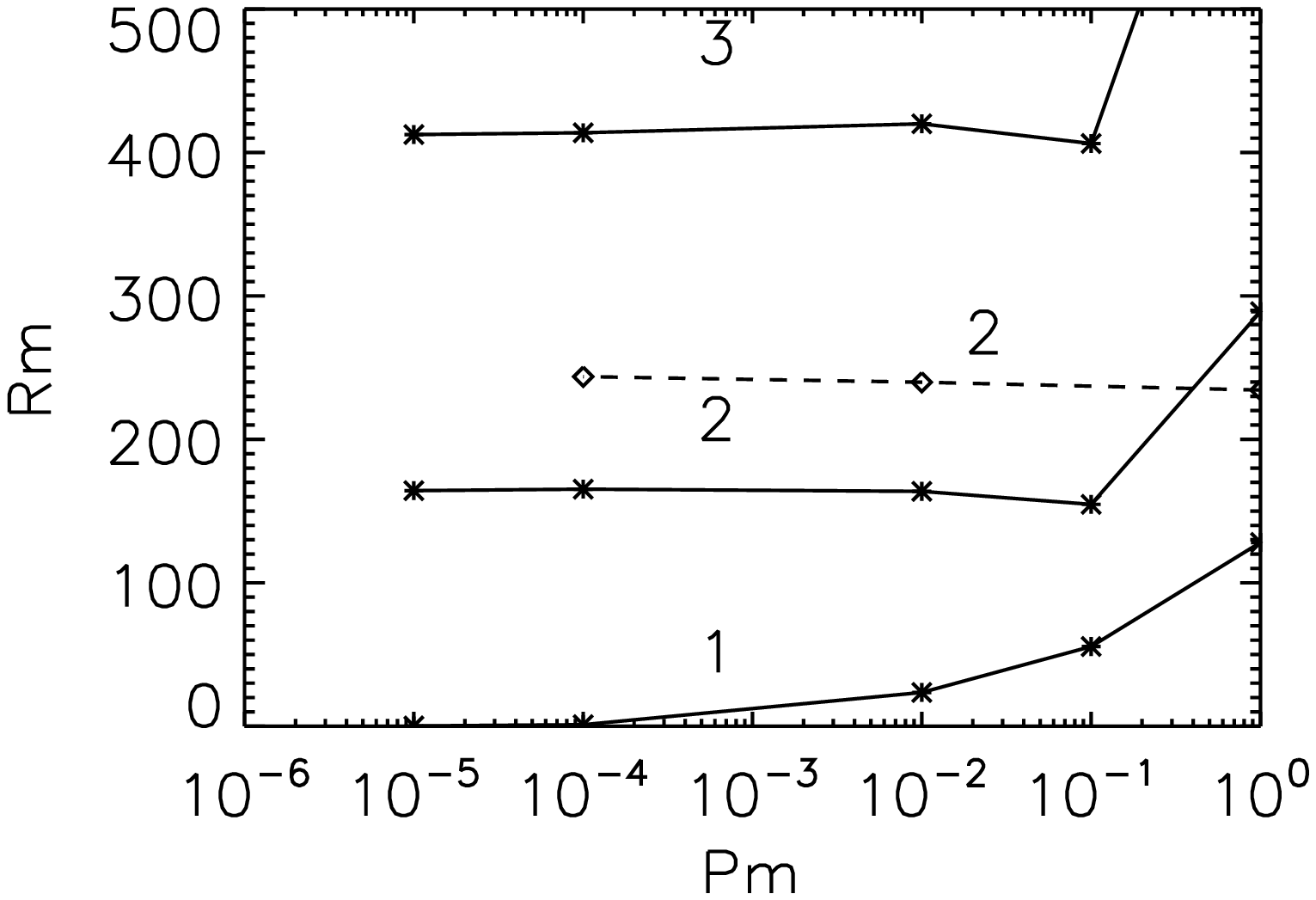}
\includegraphics[width=0.22\textwidth]{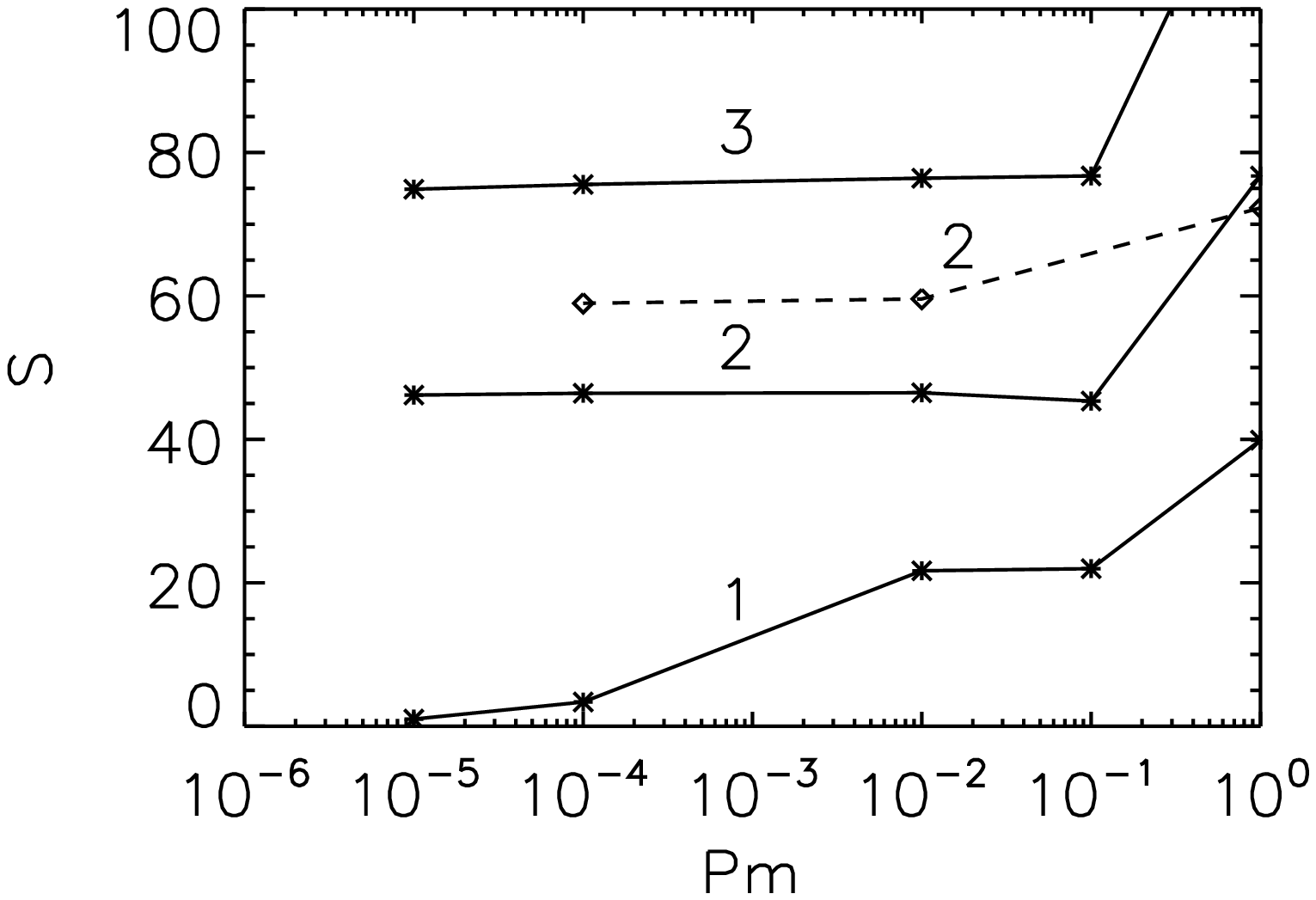}
\caption{The coordinates $\Rm$ (left) and $\S$ (right) for the minima  of the lines of
         marginal instability given in Fig. \ref{35a}. For small $\Pm$ the lines for
         $m=1$ scale with $\Re$ and $\Ha$, but for $m=2,3$ they scale with $\Rm$ and
         $\S$. The dashed lines are for the quasi-galactic rotation profile $\mu=0.5$,
         and indicate that this behaves much the same as $\mu=0.35$.}
\label{35b}
\end{figure}

Figs. \ref{35a} and \ref{35b} show the equivalents of Figs. \ref{25a} and
\ref{25b}. While there are some similarities, there are also many differences.
Most importantly, as seen in Fig. \ref{35b}, it is only for $m=1$ that the
critical parameters are $\Re$ and $\Ha$. For $m=2,3$ the instabilities
instead scale with $\Rm$ and $\S$. These new scalings as $\Pm\to0$ suggest
that these instabilities may have astrophysical applications, where
$\Pm<1$ and $\Mm>1$ are often both satisfied. Because of their scaling with
$\Rm $ and $\S$ these $m=2,3$ modes should also exist for vanishing
viscosity, $\nu=0$. They cannot be reproduced, therefore, with codes based on
the inductionless approximation ($\Pm=0$).

%%%%%%%%%%%%%%%%%%%%%%%%%%%%%%%%
\section{Kinetic and magnetic energies}
%%%%%%%%%%%%%%%%%%%%%%%%%%%%%
The kinetic and magnetic energies of magnetohydrodynamic turbulence
are often assumed to be equipartitioned. To probe this idea the ratio
\begin{equation}
 \varepsilon=\frac{\langle \vec{b}^2\rangle}{\mu_0\rho \langle \vec{u}^2\rangle}
 \label{ratio}
\end{equation}
of the two energies is calculated, averaged over the container. The stationary
background solutions (\ref{basic}) are excluded.

\begin{figure}[htb]
\centering
\includegraphics[width=0.42\textwidth]{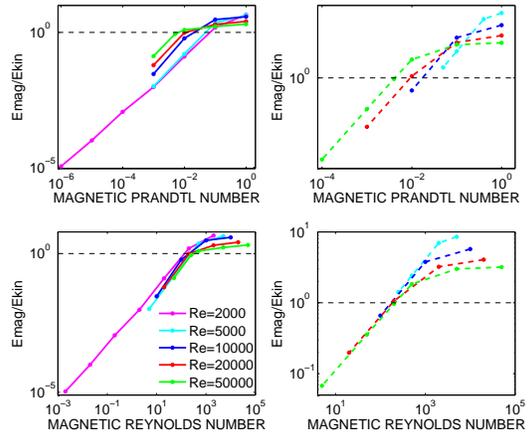}
\caption{The ratio $\varepsilon=E_{\rm mag}/E_{\rm kin}$ between magnetic and kinetic energy as a function of $\Pm$ (top)
         and $\Rm$ (bottom) for $\mu=0.25$ (left, with $\Ha=600$) and $\mu=0.35$ (right, with $\Ha=1000$).
         $\varepsilon$ exceeds unity for $\Rm\simeq 200$. $\Pm$ is not an
appropriate measure, but $\Rm$ is.}
\label{25d}
\end{figure}

In the top panels of Fig. \ref{25d} this ratio is plotted for
various Reynolds numbers as a function of the magnetic Prandtl number.
The Hartmann number is fixed, and $\mu$ takes the two values 0.25 and 0.35.
The result is that for small magnetic Prandtl number ($\Pm\lesssim 10^{-2}$)
the relation $\varepsilon\propto\Pm$ seems to hold, which implies that
$\eta \langle \vec{b}^2\rangle/\mu_0\rho\simeq \nu  \langle {\vec u}^2\rangle$,
or equivalently $b_{\rm rms} =$O$(\sqrt{\Pm} u_{\rm rms})$. This dependence is
weaker than that used by \citet{R64}, who suggested that for small $\Pm$
 $b_{\rm rms} =$O$({\Pm} u_{\rm rms})$.
For the given Reynolds numbers up to 50000, and magnetic Prandtl numbers
smaller than a {\em critical} value of (say) 0.01, the instability pattern is 
always dominated by the kinetic fluctuations. However, the critical $\Pm$ 
depends on the applied Reynolds number; it becomes smaller for increasing
$\Re$, and is evidently not the most appropriate measure to decide whether
the state is magnetically or kinetically dominated.

The plot also shows that the influence of the global Reynolds number on this
relation is only weak. For faster  rotation the ratio (\ref{ratio}) is somewhat
larger  than for slower rotation. For forced MHD turbulence models \citep{B14}
found a similar behavior for the viscous and ohmic
dissipation, but for such models the magnetic energy reservoir is only filled by
the work of the Lorentz force against the driven velocity field.
 
In the bottom panels of Fig. \ref{25d} the ratio $\varepsilon$ 
is plotted now as a function of the magnetic Reynolds number $\Rm$.
One finds a clear scaling of the curves with
$\Rm$ for both the potential rotation law $\mu=0.25$ as
well as the quasi-Keplerian law $\mu=0.35$. The magnetic energy exceeds the
kinetic energy for all $\Rm\gtrsim 200$. This behavior does not depend on the
electric current associated with the basic state (\ref{basic}).
For smaller magnetic Reynolds numbers the MHD instability  is
always dominated by the fluid motions. For larger $\Rm$ the energy ratio seems
to become constant, in agreement with \citet{R14}. Calculations with $\Rm<200$
are only weakly magnetized, while for larger $\Rm$ the pattern is
magnetically dominated. If the curves do
scale with $\Rm$ rather than $\Pm$, then fluids with $\Pm\ll 1$ will also
become magnetically dominated once $\Re$ and $\Rm$ are sufficiently large, which
is indeed the case for many astrophysical applications. This would  not be
possible if they  scaled with $\Pm$. Experiments with liquid metals as the fluid
between the cylinders will always lead to $\varepsilon<1$ unless the Reynolds
number exceeds $10^7$.

\begin{figure}[htb]
\centering
\includegraphics[width=0.45\textwidth]{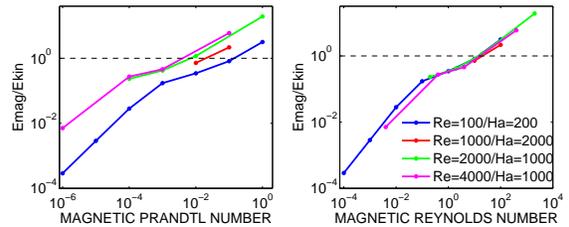}
\caption{Energy ratio $\varepsilon=E_{\rm mag}/E_{\rm kin}$ for uniform current and rigid 
         rotation, i.e. $\mu=1$. Left: for fixed $\Pm$ higher magnetic Mach numbers produce 
         higher values of $\varepsilon$. Right: the scaling with $\Rm$ is rather clear.}
\label{100b}
\end{figure}

\begin{figure}[htb]
\centering
\includegraphics[width=0.42\textwidth]{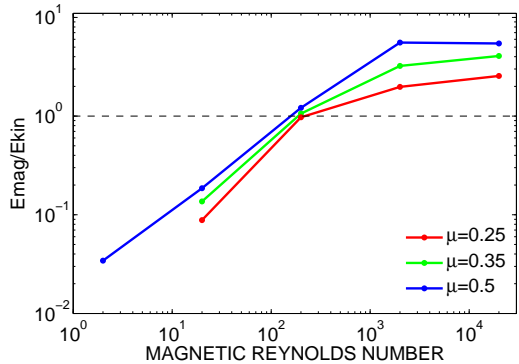}
\caption{The ratio $\varepsilon$ between magnetic and kinetic energy as a function of $\Rm$ and $\mu$ for
         $\mu=0.25/0.35/0.5$. Reynolds number is $\Re=20000$ for all three configurations, Hartmann number 
         $\Ha=600$ for $\mu=0.25$, otherwise $\Ha=1000$.}
\label{fig_epsilon_05_07_1}
\end{figure}

The energy ratio for $\mu=1$ (TI) is shown in Fig. \ref{100b}, and exhibits
the same $\Rm$-dependent characteristics. It is thus the magnetic Reynolds
number rather  than the magnetic Prandtl number which determines the
relationship of the two energies according to
\begin{equation}
 \varepsilon\propto \Rm.
 \label{ratiorm}
\end{equation}
For $\mu=1$ magnetic fields dominate for {\em critical} magnetic Reynolds
numbers of $\Rm\simeq20$ and above, roughly a factor of 10 less than for AMRI.
Fig. \ref{fig_epsilon_05_07_1} shows that even $\mu$ as large as 0.5 still
yields the previous result $\Rm\simeq200$ as the critical value. Any differential
rotation at all therefore seems to yield a much larger critical value than the
no differential rotation case $\mu=1$. From an astrophysical point of view the
distinction between $\Rm\simeq20$ and 200 is of course hardly important; most
magnetized objects are likely to have values far greater anyway. From the point
of view of laboratory experiments though a reduction in $\Rm$ by a factor of 10
could be of considerable interest.

The results in Fig. \ref{100b} not only scan over $\Pm$, but do so for various
choices of $\Ha$ and $\Re$. Converting to $\Mm$, the main result of this plot is
that the ratio $\varepsilon$ grows for increasing  $\Mm$. Hence, a pinch-type
instability for fixed magnetic Prandtl number is the more magnetic the
{\em weaker} the magnetic background field is compared with the basic
rotation rate.

%%%%%%%%%%%%%%%%%%%%%%
\section{The spectra}
%%%%%%%%%%%%%%%%%%%%%%%%%%%%%%%%%%%%%%%%%%%%%%
\subsection{Azimuthal direction}
\begin{figure}[htb]
 \centering
 \includegraphics[width=0.42\textwidth]{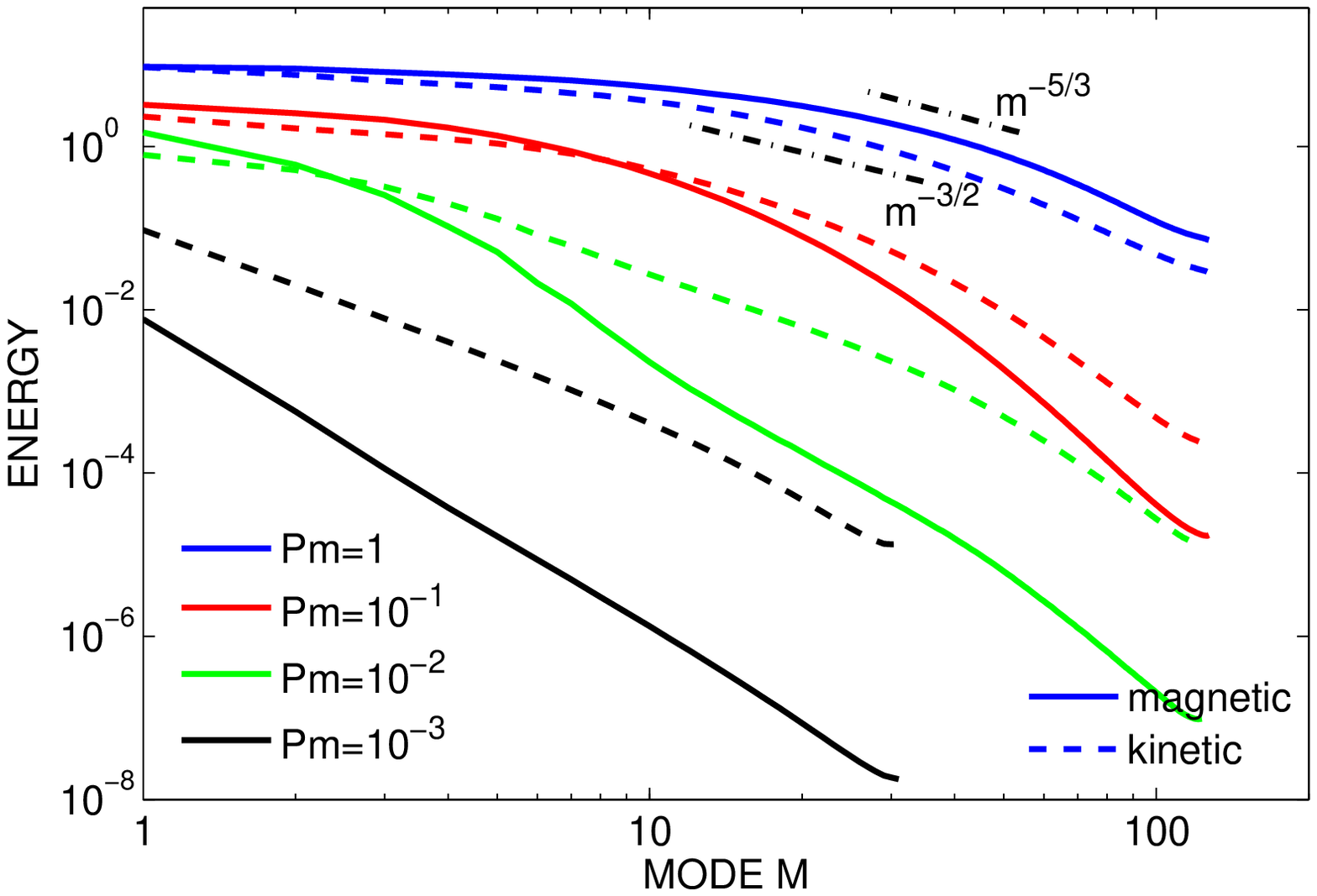}
 \includegraphics[width=0.42\textwidth]{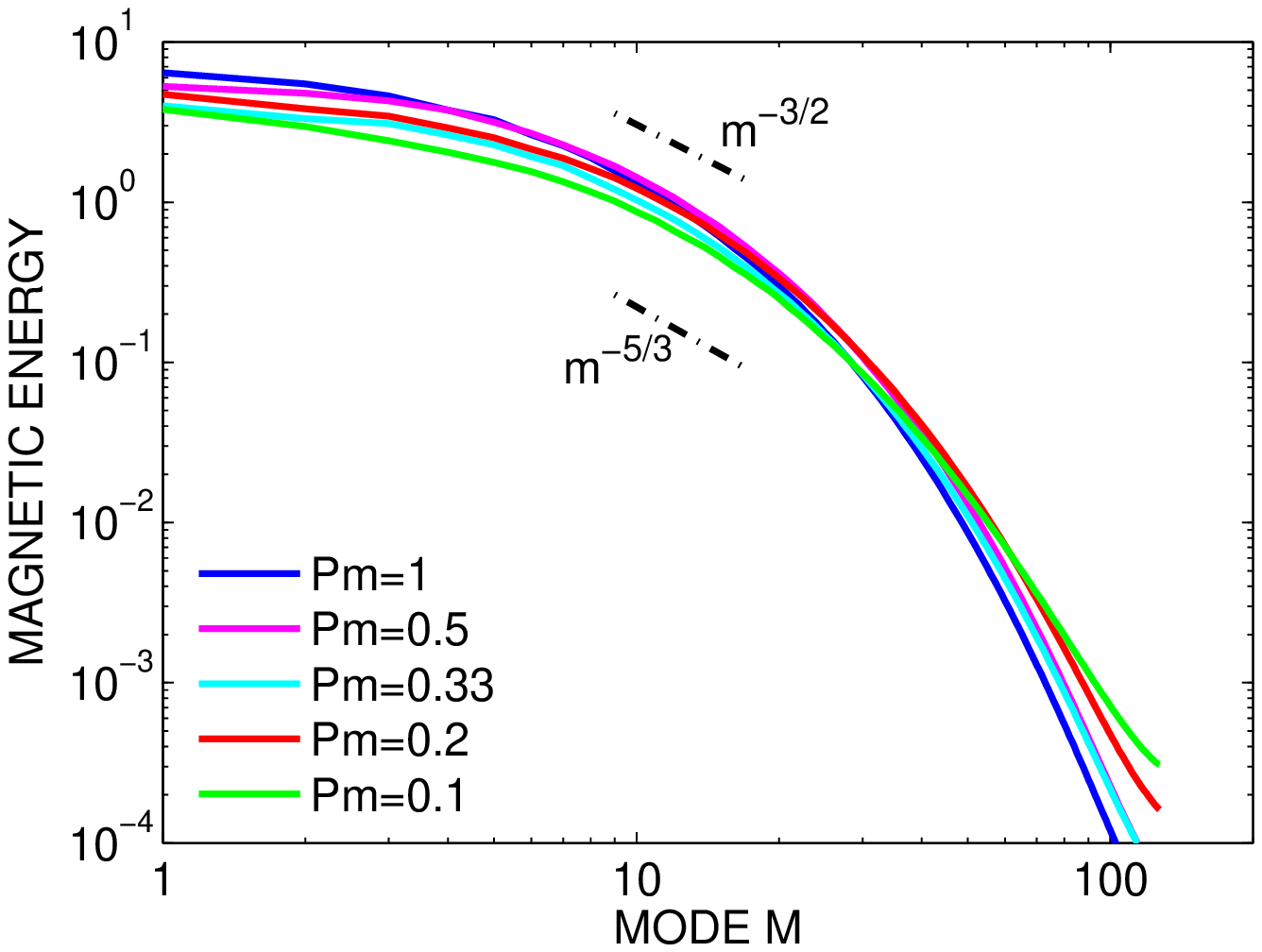}
 \caption{The spectra of the standard AMRI, for various $\Pm$.
 Top: The magnetic (solid lines) and the kinetic (dashed lines) energies
 in the azimuthal Fourier modes $m$ for  $\Re=50000$.
          Bottom: The magnetic spectra for $\Rm=10000$.
          $\Ha=600$,  $\mu=0.25$. The dashed-dotted lines represent
          the Kolmogorov spectrum and the 
          magnetohydrodynamic IK spectrum.}
 \label{spec25}
\end{figure}

It is typical for the magnetic instability under consideration that i)  only
nonaxisymmetric modes  and ii) only the  modes with the lowest  $m\neq 0$
become unstable for finite $\Ha$ and $\Re$. The rotating pinch gives an example
where only a single linearly unstable mode ($m=1$) injects the energy into the system,
where the nonlinear interactions transport it to the higher modes. In
contrast, for the standard AMRI with $\mu=0.25$ modes with higher $m$ also
become unstable if, for a given magnetic field, the system rotates fast enough but
not too fast. Figure \ref{25a} shows that for given $\Ha$ and $\Re$ the number
of unstable modes decreases for decreasing magnetic Prandtl number. This is a
consequence of the fact that for AMRI all azimuthal modes scale with $\Re$ and
$\Ha$ for $\Pm\to 0$. As a consequence, for fixed Reynolds and Hartmann numbers 
one would expect a spectrum that becomes steeper and steeper already on the
large scales (low $m$) with decreasing $\Pm$. Figure \ref{spec25} (top) shows 
the kinetic and magnetic energies for all modes $m$ for this situation of a 
fixed magnetic field with $\Ha=600$, and the very high Reynolds number of
$\Re=50000$ and several $\Pm$. The magnetic and the kinetic spectra have a 
similar shape, but they are only close together for large $\Pm$. For small $\Pm$ 
the magnetic spectrum lies below the kinetic one, as already demonstrated by 
Fig. \ref{25d}. For $\Pm$ of order unity the spectrum is rather flat (see the 
blue line corresponding to $\Rm=50000$) on the low $m$ side, and rather 
steep for small $\Pm$, where only one unstable mode exists.
 
\begin{figure}[htb]
\centering
\includegraphics[width=0.42\textwidth]{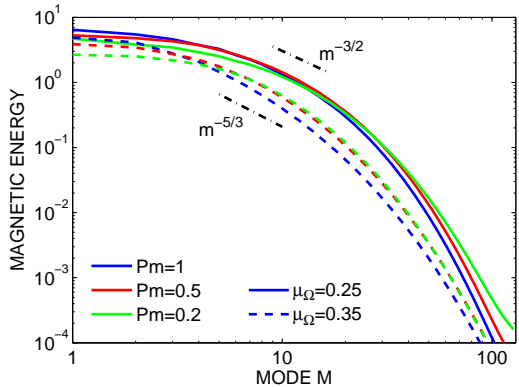}
\caption{Magnetic energy spectrum for $\Rm=10000$,
 comparison of $\mu=0.25$ and $\mu=0.35$.}
\label{spec_25_35}
\end{figure}

Magnetic spectra of AMRI for a constant magnetic Reynolds number are shown
in the bottom panel of Fig. \ref{spec25}. In this representation
the results do {\em not} depend on the magnetic Prandtl number. That is, 
large Reynolds numbers and small $\Pm$ lead to the same spectra as small Reynolds 
numbers and large $\Pm$. The combination of both panels indicates that the
spectra become increasingly flat for increasing $\Rm$. The same is true for
$\mu=0.35$, as shown in Fig. \ref{spec_25_35} for fixed $\Rm=10000$.
The comparison between the spectra of the potential flow $\mu=0.25$ and the
quasi-Keplerian $\mu=0.35$ reveals not much difference at the same $\Rm$. 
The tails of the spectra become slightly less steep
for flatter rotation profiles; the smoothing action of the differential rotation
is reduced. The scaling in the intermediate range and large scales is the same;
the total amount of magnetic energy in the quasi-Keplerian profile is reduced. 

It is also obvious that the spectra for the kinetic and
magnetic fluctuations have similar shapes, and only suggestively show a 
plateau in the intermediate $m$-range. If a power law is fitted, both 
would slightly favor the Iroshnikov-Kraichnan (IK) spectrum with $m^{-3/2}$ 
compared to the Kolmogorov spectrum $m^{-5/3}$, but the differences are 
small and not significant. Although the IK profile is favored for MHD 
turbulence \citep{Z04,Mas08}, Kolmogorov-like spectra are also known from the 
measurements of turbulence in the solar wind \citep{Mar03} as well as the result 
of 3D MHD simulations \citep{Mue00}. Often, however, the direct numerical
simulations are done for equipartition ($\varepsilon=1$) and for $\Pm$ of order
unity (see \citet{B14}). One conclusion here could be that this assumption is
reasonable if $\Rm$ is large enough. A clear preference between IK and
Kolmogorov scaling cannot be made.

\begin{figure}[htb]
\centering
\includegraphics[width=0.42\textwidth]{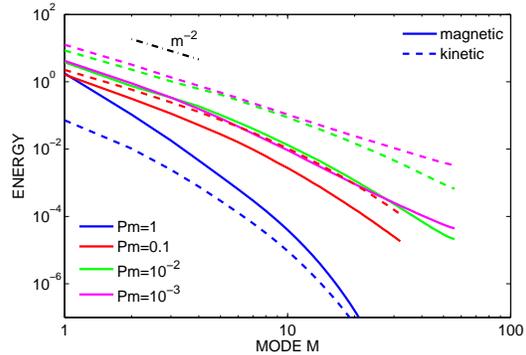}
\caption{The spectrum for the rigidly rotating pinch.
         The magnetic (solid lines)  and the kinetic (dashed lines) 
         energies in the azimuthal Fourier modes $m$ for $\Re= 2000$ 
         and $\Ha=1000$ for various $\Pm$.}
\label{spec100}
\end{figure}

We next return to the question whether the spectra are modified by the number of 
linearly unstable modes or not. As demonstrated in section \ref{subsec_pinch}, 
for the rigidly rotating pinch only $m=1$ becomes unstable. Figure \ref{spec100} 
shows the power spectra for this profile for fixed Reynolds and Hartmann number 
but various magnetic Prandtl numbers. The Mach number varies 
between $\Mm=0.2$ for $\Pm=0.01$  and $\Mm=2$ for $\Pm=1$. Only the mode $m=1$ 
provides the energy to initiate the nonlinear cascade; it is also always $m=1$ that
contains the most energy. As expected, the TI spectrum is much steeper than the 
AMRI spectrum. It is even so steep that neither the IK nor the Kolmogorov 
spectrum fit the resulting curves. Much closer comes a scaling $m^{-2}$ that is
found in forced turbulence \citep{da16} or in spectra of not yet truly turbulent 
flows \citep{wa16}. Because the AMRI power spectra for low $\Rm$ also have a 
tendency towards $m^{-2}$, this might be a sign of very weak turbulence.

On the other hand, as the energy source in this case is only from the underlying
current rather than any differential rotation, one might question whether $\Re$
and/or $\Rm$ are the relevant measures at all, or whether $\Ha$ might not be
the more appropriate measure in determining the shape of the spectrum for the
rigidly rotating pinch. The largest numerically accessible Hartmann number is
$\Ha\approx2000$, and still showed no deviation from this $m^{-2}$ scaling.

%%%%%%%%%%%%%%%%%%%%%%%%%%%%%%%%%%%%
\subsection{Axial direction}
\begin{figure}[htb]
\centering
\includegraphics[width=0.42\textwidth]{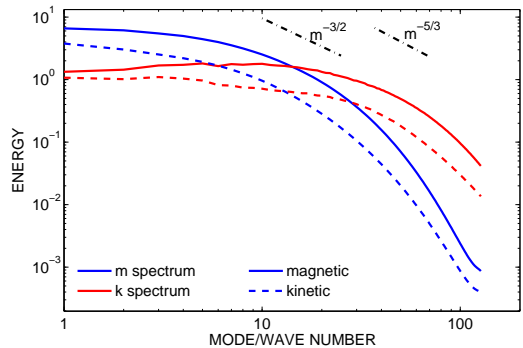}
\caption{Energy spectra of $m$ and $k$ wavenumbers for
 $\Rm=20000, \Ha=600, \Pm=1, \mu=0.25$.}
\label{fig_spec_km_ray}
\end{figure}

The spectra in the axial direction have a somewhat different shape compared
with the azimuthal direction.
The basic wavenumber at the onset of instability is $k\approx4-5$,
corresponding to a round cross section of the patterns.
A small increase in the Reynolds number extends this range to 
$2\lesssim k\lesssim 8$. For turbulence at even higher Reynolds numbers, these
large scales remain as a plateau for $k\lesssim 8$, and the part of the spectra for intermediate $k$ shows a similar
behavior as the $m$ spectra with no significant plateau
(Fig. \ref{fig_spec_km_ray}). The closest slope is again the 
IK profile with $k^{-3/2}$. 

\begin{figure}[htb]
\centering
\includegraphics[width=0.42\textwidth]{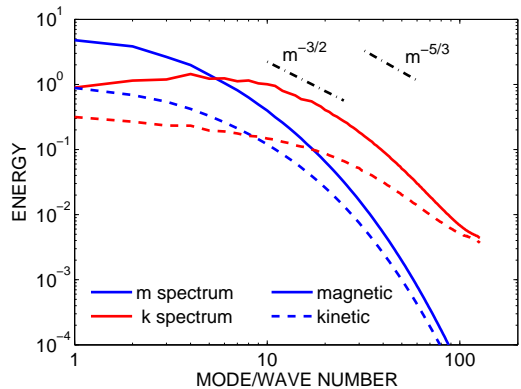}
\caption{Energy spectra of $m$ and $k$ wavenumbers for
 $\Rm=10000, \Ha=1000, \Pm=1, \mu=0.35$.}
\label{fig_spec_km_kep}
\end{figure}

One aspect where the $m$ and $k$ spectra clearly differ is for large values.
As previously noted, for large $m$ the spectra drop off quite strongly, due
to the smoothing and hence damping effect of the differential rotation.
Such a mechanism does not exist in the axial direction, and larger $k$ are
correspondingly more strongly excited than large $m$. The greater the shear,
the greater the difference between $m$ and $k$ in this regard. For
$Rm\approx 20000$ the largest $k$ are stronger by one order of magnitude
for the quasi-Keplerian flow, and two orders of magnitude for the steeper
potential flow (see Fig. \ref{fig_spec_km_ray} and \ref{fig_spec_km_kep}).
For the very large $\Rm$ of real astrophysical objects, this anisotropy
between different directions might be even more strongly developed.

%%%%%%%%%%%%%%%%%%%%%%%%%
\section{Summary} 
%%%%%%%%%%%%%%%%%%%%%%%%%
Magnetohydrodynamic Taylor-Couette flows have been investigated for many
decades \citep{R64}. One possibility that is always stable for ideal flows is
if the imposed field is purely azimuthal, and has the same radial profile as
the imposed velocity profile \citep{C56}. However, as demonstrated by
\citet{R15b}, such Chandrasekhar states can become unstable if at least one
of the diffusivities is non-zero. If viscosity and magnetic resistivity are
{\em both} non-zero, the $m=1$ marginal instability curves in the $\Ha$-$\Re$
plane become independent of magnetic Prandtl number in the limit $\Pm\to 0$.
From the definition (\ref{Mm}), there will then always exist some (small)
value of $\Pm$ below which {\em all} eigenvalues of the linear perturbation
equations yield $\Mm<1$. Given that many cosmical objects such as accretion
disks, stars and compact objects often combine small $\Pm$ and large $\Mm$,
the stability of these Chandrasekhar states might therefore seem to be a
purely academic exercise. This is especially the case as for the standard
AMRI at least, with $\Om\propto 1/R^2$ and $B_\phi\propto 1/R$, the $m>1$
modes exhibit exactly the same scaling with $\Ha$ and $\Re$. Similarly, for
the pure TI, with $\Om=\rm const$ and $B_\phi\propto R$, only the $m=1$
mode is unstable, and it also scales with $\Ha$ and $\Re$ for small $\Pm$.

However, as we demonstrated in this work, for rotation laws {\em between}
$\Om\propto 1/R^2$ and $\Om=\rm const$ (and corresponding $B_\phi\propto R
\Om$), the $m>1$ modes behave differently, for small $\Pm$ scaling instead
with $\S$ and $\Rm$.  From the basic relationship $\Mm=\Rm/\S$, together
with the upward-sloping shape of the critical stability curve $\Rm=\Rm(\S)$,
it then follows that $\Mm>1$ can always be achieved, even in the limit
$\Pm\to0$. This finding is one of the main conclusions of this work, and
suggests that only the $m>1$ modes are relevant for the majority of
astrophysical applications.
  
The magnetic and kinetic energies of MHD instabilities are often considered
as approximately the same order when $\Pm\simeq 1$. For smaller $\Pm$ the
magnetic energy is assumed to be smaller than the kinetic energy \citep{R64}.
This is indeed true for these Chandrasekhar states. The top panels
of Figs. \ref{25d} (AMRI) and \ref{100b} (TI) show the ratio (\ref{ratio})
for the two limiting examples for various $\Pm$. In
both cases the magnetic and kinetic energies are indeed equipartitioned for
$\Pm\simeq 1$, and $\varepsilon\ll 1$ for smaller $\Pm$. For the  curves
with fixed $\Ha$ and $\Re$  a clear trend exists  of the critical $\Pm$ at the
crossing points at the axis $\varepsilon=1$. For a single curve for the pair
$[\Ha,\Re]$  the ratio scales as $\varepsilon\propto \Pm$, but the curves
with other parameter combinations are not identical but rather parallel.  
 
As this conclusion holds for both an example with differential rotation
(AMRI) and another one with rigid rotation (TI), the induction by the
background flow is obviously not so important. Moreover it is not $\Pm$ that
defines the value of $\varepsilon$. The relevant parameter is the magnetic
Reynolds number. This is true not only for the limits $\mu=0.25$ and $\mu=1$,
but also for all $\mu$ in between.
  
For non-magnetic Taylor-Couette flows \citet{D07} simulated turbulent
solutions  with $r_{\rm in}=0.5$ for flows with resting outer
cylinder. The critical Reynolds number for $m=0$ is 68, for $m=1$ it is 75, and
for $m=2$ it is 127 \citep{R67}. For $\Re=1000$ the flow is not yet turbulent
as no high frequencies appear. For $\Re=3000$, 5000 and 8000 temporal
power spectra of the  Kolmogorov-type develop, which only differ slightly for
high frequencies. The higher the Reynolds number the higher frequencies appear
as more and more nonaxisymmetric modes become unstable. A similar behavior can
be observed for the AMRI $m$ spectra of Fig. \ref{spec25}. The bottom panel
displays spectra of the magnetic energy for Reynolds numbers from $\Re=10^4$
(blue line) to $\Re=10^5$ (black line). The latter line represents the
occurrence of higher frequencies.

The top panel of Fig. \ref{spec25} demonstrates the influence of the magnetic
Prandtl number for given Hartmann and Reynolds numbers, in comparison to
the results of Fig. \ref{25a}. The majority of the modes are unstable for
$\Pm=1$, while for smaller $\Pm$ (or more general $\Rm$) the higher modes become 
more and more stable so that the steepest curve in Fig. \ref{spec25} (top) results 
for the smallest $\Pm$. 

The opposite is true for the azimuthal power spectrum of the rigidly-rotating 
pinch. According to Fig. \ref{spec100} the curve for $\Pm=1$ is the steepest.
Here only the mode with $m=1$ is unstable, with the strongest rotational
suppression for $\Pm=1$ (see Fig. \ref{100a}). Even the power spectrum of the
rigidly-rotating pinch gives an indication about the double-diffusive character
of the nonaxisymmetric magnetic kink-type instability, as analyzed in detail
by \citet{R16}.

The scaling behavior of the intermediate range of both wavenumbers $m$ and $k$ remains 
unclear in the sense that no significant plateau develops. The closest scaling exponent will
be $m^{-3/2}$ and $k^{-3/2}$ of a Iroshnikov-Kraichnan spectrum.
Kolmogorov's $-5/3$ scaling is not observed. 

In comparison to the azimuthal spectra, the axial spectra  show a 
different distribution. First of all there exists a large-scale plateau
around the marginal unstable wavenumber $k=4$. The largest wavenumbers are
also much more strongly excited. The reason is the smoothing action of differential rotation, 
which tends to destroy high wavenumbers and leads to
a steeper slope in the tails of the $m$ spectra compared with the $k$ spectra.
This anisotropy should be even more strongly pronounced for the very large
$\Rm$ of real astrophysical objects.

\acknowledgments 
This work was supported by the framework of the Helmholtz Alliance LIMTECH.

%%%%%%%%%%%%%%%%%%%%%%%%%%%%%%%%%%%%%%%%%%%%%%%%%%%%%%%%%%%%%%%%%%%%%%%%%%%%%%%
\end{document}